\documentclass[fleqn,usenatbib]{mnras}

\usepackage{newtxtext,newtxmath}

\usepackage[T1]{fontenc}

\DeclareRobustCommand{\VAN}[3]{#2}
\let\VANthebibliography\thebibliography
\def\thebibliography{\DeclareRobustCommand{\VAN}[3]{##3}\VANthebibliography}

\usepackage{soul}

\usepackage{graphicx}	
\usepackage{amsmath}	
\usepackage{xcolor} 

\title[Impact of the magnetic field configuration]{Nucleosynthesis in magnetorotational supernovae: impact of the magnetic field configuration}

\author[M. Reichert et al.]{
M. Reichert$^{1}$\thanks{E-mail: moritz.reichert@uv.es},
M. Bugli,$^{2,3,4}$
J. Guilet,$^{3}$
M. Obergaulinger,$^{1}$
M. \'A. Aloy,$^{1,5}$
and A. Arcones,$^{6,7}$
\\
$^{1}$Departament d'Astonomia i Astrof\'{\i}sica, Universitat de Val\`encia, C/Dr. Moliner, 50, 46100 Burjassot (Val\`encia), Spain\\
$^{2}$Dipartimento di Fisica, Università di Torino, Via P. Giuria 1, I-10125 Torino, Italy\\
$^{3}$Universit\'e Paris-Saclay, Universit\'e Paris Cit\'e, CEA, CNRS, AIM, F-91191, Gif-sur-Yvette, France\\
$^{4}$INFN, Sezione di Torino, Via Pietro Giuria 1, I-10125 Torino, Italy\\
$^{5}$Observatori Astronòmic, Universitat de València, 46980 Paterna, Spain\\
$^{6}$Institut für Kernphysik, Technische Universität Darmstadt, Schlossgartenstr. 2, 64289 Darmstadt, Germany\\
$^{7}$GSI Helmholtzzentrum f\"ur Schwerionenforschung GmbH, Planckstr. 1, Darmstadt 64291, Germany
}

\date{Accepted XXX. Received YYY; in original form ZZZ}

\pubyear{2015}

\begin{document}
\label{firstpage}
\pagerange{\pageref{firstpage}--\pageref{lastpage}}
\maketitle

\begin{abstract}
The production of heavy elements is one of the main by-products of the explosive end of massive stars. A long sought goal is finding differentiated patterns in the nucleosynthesis yields, which could permit identifying a number of properties of the explosive core. Among them, the traces of the magnetic field topology are particularly important for \emph{extreme} supernova explosions, most likely hosted by magnetorotational effects. We investigate the nucleosynthesis of five state-of-the-art magnetohydrodynamic models with fast rotation that  have been previously calculated in full 3D and that involve an accurate neutrino transport (M1). One of the models does not contain any magnetic field and synthesizes elements around the iron group, in agreement with other CC-SNe models in literature. All other models host a strong magnetic field of the same intensity, but with different topology. For the first time, we investigate the nucleosynthesis of MR-SNe models with a quadrupolar magnetic field and a 90 degree tilted dipole. We obtain a large variety of ejecta compositions reaching from iron nuclei to nuclei up to the third r-process peak. We assess the robustness of our results by considering the impact of different nuclear physics uncertainties such as different nuclear masses, $\beta^{-}$-decays and $\beta^{-}$-delayed neutron emission probabilities, neutrino reactions, fission, and a feedback of nuclear energy on the temperature. We find that the qualitative results do not change with different nuclear physics input. The properties of the explosion dynamics and the magnetic field configuration are the dominant factors determining the ejecta composition.  
\end{abstract}



\section{Introduction}
With the detection of the gravitational wave event GW170817 \citep[][]{Abbott2017} and the following electromagnetic counterpart on the 17th of August 2017, our understanding of the synthesis of heavy elements within our universe has made a major step. It became clear that neutron star mergers exist and produce heavy elements such as Strontium \citep{Watson2019} or lanthanides \citep[e.g.,][]{Tanvir_2017ApJ...848L..27}. However, even though observed, some riddles remain and do not fully add up. The only detected event was located in a galaxy without active star formation. How is it then possible that we observe heavy elements in the atmosphere of stars such as our Sun? Are neutron star mergers already acting early enough to contribute to the composition of very old stars? Can neutron star mergers also be responsible for stars with a limited amount of heavy elements? Questions like this are common and repeating topics of current research (e.g., \citealt{Cote2019,Simonetti2019,Kobayashi2023}; see \citealt{Cowan2021} and \citealt{Arcones2023} for recent reviews on the production of elements).

An additional, rare, cosmological event that acts at early times in the galactic history would help to clear most of the remaining questions \citep[e.g.,][]{Argast2004,Matteucci2014a, Wehmeyer2015,Schoenrich2019,Siegel2019,Cote2019,Simonetti2019,Kobayashi2020,Skuladottir2020,Reichert2020,Reichert2021b,Molero2021,Molero2023}. For this event several candidates have been proposed. Among them are collapsars \citep[][]{MacFadyen_1999ApJ...524..262,Pruet2003,Surman2004,McLaughlin2005,Fujimoto2008,Siegel2019,Miller2020,Zenati2020,Barnes2022,Just2022b}, common envelope jet supernovae \citep[][]{Soker2013,Papish2014}, and magnetorotationally driven core-collapse supernovae \citep[MR-SNe,][]{Burrows2007,Takiwaki2009,Moesta2015,ObergaulingerAloy2017,Mueller2020,Kuroda2020,Obergaulinger2021,Obergaulinger2022,Bugli2020,Bugli2021,Bugli2023,Matsumoto2022,Varma2022,Powell2022}. 

For every of the aforementioned events, it is still not theoretically understood whether very heavy elements (so called r-process elements) can be produced. Current state-of-the-art simulations provide contradictory results, and more investigations with less approximate physics inputs are necessary \citep[see][for an overview]{Obergaulinger2020_handbook_nuclear_physics}. MR-SNe have a very lively past of claims in favour and against the production of r-process elements. Initially, \citet[][]{Nishimura2006}, \citet[][]{Winteler2012}, and \citet[][]{Nishimura2015} claimed that they are viable candidates with the restriction that the magnetic field has to be sufficiently strong. However, the neutrino luminosities were taken into account only approximately and later \citet[][]{Nishimura2017} showed that the combination of the neutrino properties as well as the magnetic field strength are critical for the evaluation of this question. However, also in \citet[][]{Nishimura2017}, neutrinos were only parametrized because a detailed neutrino transport was computationally too expensive. This gap was closed by \citet[][]{Reichert2021a} who calculated the nucleosynthesis of models of \citet[][]{ObergaulingerAloy2017} that included a sophisticated neutrino transport. The result was that for strong magnetic field strengths, indeed heavy elements can be synthesized, but some caveats remain as the models do not reproduce the observation of the heaviest known elements, the actinides. Furthermore, all simulations were based on 2D axisymmetric assumptions, an approximation that was shown by \citet[][]{Moesta2018} to significantly impact the neutron-richness of the models and therefore the ability to synthesize heavy elements. Recently \citet[][]{Reichert2023} showed that heavy elements can also be synthesized in 3D models, but in a reduced amount. They showed that not only the neutron-richness, but also the entropy may play a role in these events. The latter was observed in models with a radial profile of the magnetic field that was different from the usual assumed large-scale dipole. In general, most of previous nucleosynthesis studies have considered large-scale dipoles only. The magnetic field configuration stays unexplored. Exploring the field configuration is also motivated by the recent observations of the neutron star PSR J0030+0451 by the Neutron Star Interior Composition Explorer (NICER, \citealt{Nicer}) that showed that the magnetic field topology can be much more complex than a large scale dipole \citep{Bilous2019}. Furthermore, the orientation of the magnetic field plays a role in the explosion dynamics. This has been shown by \citet[][]{Halevi2018} who studied models with a 15, 30, and 45 degree tilted dipole. The most inclined dipolar configurations disfavoured the production of heavy elements. However, \cite{Halevi2018} only used an approximate neutrino leakage scheme, along with a parametrized neutrino luminosity, which limits the reach of their conclusions.

Another riddle connected to MR-SNe is their role in the detection of hypernovae. Are MR-SNe energetic enough to be classified as such? Connected nucleosynthetic fingerprints of hypernovae are Zn and $^{56}$Ni that powers the light-curve of the event. \citet[][]{Nishimura2017} pointed out that there is a balance between having neutron-rich ejecta (therefore being able to produce r-process elements) and ejecting large amounts of Zn or $^{56}$Ni that is produced in symmetric conditions. How much Zn or $^{56}$Ni MR-SNe are able to synthesize is also still under debate. \citet[][]{Nishimura2017} finds a low amount of $^{56}$Ni in the most neutron-rich explosions and \citet[][]{Grimmett2018} obtains a low amount of Zn for their 1D hypernovae models. In contrast, \citet[][]{Grimmett2021} and \citet[][]{Reichert2023} find a high amount of both. 

Here, we want to shed light into two kinds of uncertainties of MR-SNe nucleosynthesis, one that comes from the hydrodynamic evolution of the ejecta and one from the nuclear physics input. The uncertainty involved by nuclear physics input such as nuclear masses, beta decays, or fission arises as the r-process involves extreme neutron-rich nuclei that are not yet explored experimentally and thus their properties rely on uncertain theoretical models \cite[see e.g., ][]{Arnould2007, Horowitz2019, Cowan2021}. To study hydrodynamical uncertainties we post-process the state-of-the-art 3D models of \citet[][]{Bugli2021}, which explore the effects of the magnetic topology on the explosion dynamics. Similarly to previous studies \citep[][]{Reichert2021a,Reichert2023}, these models include a sophisticated neutrino transport (the M1 scheme, see \citealt{Just2015}) that significantly improve the accuracy of explosive nucleosynthesis calculations. To get a glimpse into the dependency on magnetic field configurations and orientations, the same strong magnetic field intensity has been employed for four models \citep[for a detailed presentation of the models see][]{Bugli2021}. To investigate the impact of a tilt in the dipole orientation, a model with a 90 degree tilted dipole is investigated. This is similar, but a more extreme case compared to the one of \citet[][]{Halevi2018}. A novelty here is the investigation of a quadrupolar field configuration. For comparison and reference, we also investigate one model without magnetic field. This supplements the models investigated in \citet[][]{Reichert2023}, who restricted their initial models to axially-symmetric magnetic topologies. Additionally, we study a large-scale aligned dipole with only slight variations compared to the one used in \citet[][]{Reichert2023}. The nucleosynthesis of all models is calculated varying a number of nuclear physics inputs: nuclear masses, $\beta^{-}$-decays, neutrino reactions, fission, and a feedback of the nuclear energy generation on the temperature. 

We discuss the magnetohydrodynamic (MHD) models and the details of the nucleosynthesis calculations in Sect.~\ref{sct:simulations}. The nucleosynthetic yield is presented in Sect.~\ref{sct:yield}, nuclear physics uncertainties in Sect.~\ref{sct:nuc_phys_unc}. A comparison to observations is shown in Sect.~\ref{sct:observations}. We close our study with a discussion and summary in Sect.~\ref{sct:conclusion}.

\section{Simulations}
\label{sct:simulations}
In this section we present the fundamental tools, inputs, and models of the magnetohydrodynamic simulations as well as the nucleosynthesis calculations.
\subsection{Neutrino-(M)HD simulations}

Here, we present the progenitor model and the neutrino-(M)HD simulations of \citet{Bugli2021}. Next, we will shortly summarize the simulation setup and dynamics of the models, whose more detailed description can be found in \citet{Bugli2020} and \citet{Bugli2021}.
\begin{table*}
 \caption{Overview of the properties of our models. The first column indicates the model name, the second the magnetic field topology, the third column the magnetic field intensity \citep[for more information see][]{Bugli2021}, the fourth column the inclination of the magnetic field relative to the rotational axis, the fifth column the treatment of gravity, the sixth column the final simulation time after bounce. The last three columns give the ejecta mass and diagnostic energy at the end of the simulation as well as the amount of traced and calculated tracer particles.
 }
 \label{tab:sim_properties}
 \begin{tabular}{lcccccccc}
  \hline
  Model 
  & Topology & B$_0$ & Inclination & Gravity
  & $t_{\rm f, pb}$ 
  & Ejected Mass & Diag. energy & \#Tracers \\
  & 
  & [G] & [$^{\circ}$]& &$[\mathrm{s}]$
  & [$M_{\sun}$] & [$10^{51}\,\mathrm{erg}$] &  Tot. / Calc.\\
  \hline
  H     & -         & $0$        & -  & A & 0.706 & 0.14 & 0.40 & 875460 / 851    \\
  L1-90 & Dipole    & $10^{12}$  & 90 & A & 0.789 & 0.27 & 0.41 & 1922732 / 1836  \\
  L2-0A & Quadrupole& $10^{12}$  &  0 & A & 0.496 & 0.41 & 1.01  & 2101476 / 1822  \\
  L2-0B & Quadrupole& $10^{12}$  &  0 & B & 0.886 & 0.33 & 0.83 & 1963280 / 1928  \\
  L1-0  & Dipole    & $10^{12}$  &  0 & A & 0.525 & 0.35 & 2.89 & 2201788 / 2651  \\
  \hline
 \end{tabular}
 
\end{table*}
The simulations are performed in full 3D with the MHD code \textsc{Aenus-Alcar} \citep{Obergaulinger2008,Just2015,ObergaulingerAloy2017} and they include an M1 scheme for the neutrino transport \citep[see][for details of the implementation]{Just2015}. All simulations use the EOS of \citet{LattimerSwesty1991} with an incompressibility of $K=220\,\mathrm{MeV}$. The angular resolution is set to $128$ and $64$ equally spaced angles in $\phi$ and $\theta$ direction, respectively. The gravitational potential includes the general relativistic corrections developed in \cite{Mareketal2006}, specifically, their function $A$ (with one exception, see below). The radius is distributed over $210$ grid points, with a uniform spacing of $0.5\,\mathrm{km}$ up to $10\,\mathrm{km}$, and a logarithmic spacing for larger radii up to $r_\mathrm{max} \approx 8.8 \times 10^4$. The stellar progenitor of all the simulations was chosen to be the same Wolf-Rayet star, i.e. model 35OC of \citet{woosley-heger2006}, having a zero age main sequence mass of $35\,\mathrm{M_\odot}$ and a collapse mass of $28.1\,\mathrm{M_\odot}$. The core of this model rotates at a frequency $\Omega_{\rm c}\approx 2\,\text{rad}\,\text{s}^{-1}$ (at $r=0$), which slightly decreases outwards ($\Omega\approx 1\,\text{rad}\,\text{s}^{-1}$ at $r\approx 10^8\,$cm). The progenitor star 35OC is identical with the one used in previous investigations \citep[][]{ObergaulingerAloy2017,Obergaulinger2020,Obergaulinger2021,Aloy2021,Bugli2020,Bugli2021,Bugli2023} and nucleosynthesis studies \citep{Reichert2021a,Reichert2023}. In total, we investigate the nucleosynthesis of four neutrino-(M)HD-simulations, where we mostly allow for variations in the topology of the magnetic field. 
This is justified by our incomplete knowledge of the magnetic field topology stemming either from pre-collapse stellar evolution or from dynamos in the PNS. All the models including magnetic fields yield supernova explosions driven (in different degrees) by magneto-rotational effects and, thus, we may catalogue them as MR-SNe.

The magnetic field in all magnetized models is constructed starting from an "unrotated" toroidal vector potential $A^\phi_l$ defined as \citep{Bugli2021} 
\begin{equation}\label{eq:Aphi}
    A^\phi_l(r,\theta)= r\frac{B_0}{(2l+1)}\frac{r_0^{3}}{r^{3}+r_0^{3}} \frac{P_{l-1}(\cos\theta)-P_{l+1}(\cos\theta)}{\sin\theta},
\end{equation}
where $B_0$ is a normalisation constant, $l$ is the order of the multipolar component, $r_0$ is a characteristic radius within which the field is roughly constant, and $P_l$ represents the Legendre polynomial of order $l$.
This vector field is then modified to account for an inclination between the rotational and magnetic axes by an angle $\alpha$ \citep{Halevi2018}, leading to radial, polar, and azimuthal components of the vector potential defined by
\begin{align}
    A_r & =  0, \\
    A_\theta & =  -r\frac{B_0}{2}\frac{r_0^{3}}{r^{3}+r_0^{3}}\sin\phi\sin\alpha,\\
    A_\phi & =  r\frac{B_0}{2}\frac{r_0^{3}}{r^{3}+r_0^{3}}(\sin\theta\cos\alpha-\cos\theta\cos\phi\sin\alpha).
\end{align}

{\textbf{Model L1-0} is initiated with a strong aligned magnetic dipole ($\alpha=0^{\circ}$). The setup is very similar to model S in \citet{Obergaulinger2021} and \citet{Reichert2023} with the difference that it does not start with a toroidal field component during collapse. This model develops the earliest (magneto-rotational) supernova explosion among the ones considered here. It develops two-sided, quasi-symmetric, bipolar jets. The supernova shock propagates faster than in any other case, reaching a radius $\approx 10^4\,$km within less than $400\,$ms after core bounce. This keeps increasing the angular momentum of the compact remnant (i.e. of the PNS) and, due to its large centrifugal support, the (oblate) compact remnant contracts at the smallest rate of all the models of this work, and has an equatorial radius of $\sim 80-90\,$km during most of the computed post bounce time. Among all presented models, this one is the one hosting the most neutron-rich conditions. The difference compared to the S model of  \citet{Reichert2023} in the initial magnetic field setup is that model S includes both an aligned dipolar magnetic field and a toroidal field in equipartition \citep[$10^{12}$ G, see Table~1 in][]{Obergaulinger2021}, while model L1-0 includes a pure magnetic dipole. This is already enough to create slightly more neutron-rich conditions in model L1-0 with a minimum value at $7\, \mathrm{GK}$ of $Y_e\sim 0.19$ compared to $Y_e\sim 0.23$ in model S. 
Model L1-0 was the only case in which we did not directly use the data published in \cite{Bugli2021}, as we noticed the presence of a spurious radial magnetic field in a small region of the progenitor that was set by accident at the beginning of the stellar collapse.
Since nucleosynthesis calculations are extremely sensitive to the evolution of the electron fraction and the thermodynamic properties of the ejecta, we decided to re-run model L1-0 with the correct initial conditions.
This new simulation produced the same prototypical magneto-rotational explosions presented in \cite{Bugli2021} (although slightly more powerful), while the resulting yields showed non-negligible differences for $A\gtrsim$150.
Therefore, in the rest of the paper we always refer to this new realisation of model L1-0.

\textbf{Model L1-90} is initiated with a magnetic dipole that is tilted by $\alpha=90^\circ$ with respect to the rotational axis (i.e., the case of a \emph{perpendicular} dipole). 
This particular configuration is motivated by the fact that MRI-driven dynamo processes taking place within the PNS after shock formation have been found to produce a highly tilted magnetic dipolar component \citep{reboul-salze2021}.
This model produces the weakest explosion and the slowest shock among the magnetized cases, as the field's topology disfavours the extraction of rotational energy through the polar caps of the PNS \citep{Bugli2021}, thus decreasing the efficiency of the magneto-rotational mechanism.
Despite the shock front being rather spherical, the model nevertheless builds up high-entropy columns of hot ejecta that propagate along the rotational axis and can be as neutron-rich as $Y_e\sim 0.27$ (at $7\,\mathrm{GK}$).

\textbf{Model L2-0A} is initiated with a quadrupolar magnetic field aligned with the progenitor's rotational axis ($\alpha=0^{\circ}$).
The relevance of high-order magnetic multipoles comes, once again, from recent studies focusing on PNS dynamos driven by MRI \citep{reboul-salze2022} and convection \citep{Raynaud2020}, which show how the in-situ amplification of magnetic fields tends to produce components with finer angular resolution than a magnetic dipole.
Since a quadrupolar magnetic field is less efficient than an aligned dipole at transferring rotational energy from the PNS to the polar outflows \citep{Bugli2020}, this model leads to a weaker supernova explosion w.r.t. L1-0. Despite the initial prompt expansion, the bipolar outflows are significantly less collimated and the shock front expands less rapidly.
However, this model produces also the slowest spinning PNS among all cases considered, as the non-vanishing radial field in the equatorial region very efficiently transports angular momentum from the PNS interior to its surroundings at low latitudes, which are not promptly ejected in the outflows.

\textbf{Model L2-0B} has the same magnetic field setup as model L2-0A and differs only by a different pseudo-Newtonian gravitational potential \citep[case $B$ of][]{Mareketal2006}. While the difference between models L2-0A and L2-0B is of technical nature, including this in our set of models allows us to assess whether different recipes to weaken the Newtonian gravitational potential (to approximate better the general relativistic one) have an impact on the nucleosynthetic yields of the supernova explosion. We note that it was pointed out in \citet{Mareketal2006} that this pseudo-Newtonian gravitational potential is a worse approximation of the general relativistic potential compared to the gravitational potential used in our other models.

\textbf{Model H} is a non-magnetised model that serves as reference for a neutrino-driven explosions aided by fast rotation.
Due to the lack of initial strong magnetic fields, the shock stalls at $\sim200\,$km for $\sim200\,$ms, after which it starts expanding again thanks to the neutrino-heating mechanism and the onset of the low $T/|W|$ instability \citep{Bugli2023}. 
The ejecta are predominantly distributed in the North hemisphere, with no development of jets (left-hand panel of Fig.~\ref{fig:ye_models}). They are dominated by mostly symmetric matter ($Y_e \sim 0.5$), however, also hosting slightly neutron-rich matter with minimum values of $Y_e \sim 0.42$ at $7\,\mathrm{GK}$.

\begin{figure*}
\begin{center}
\includegraphics[width=1.0\linewidth]{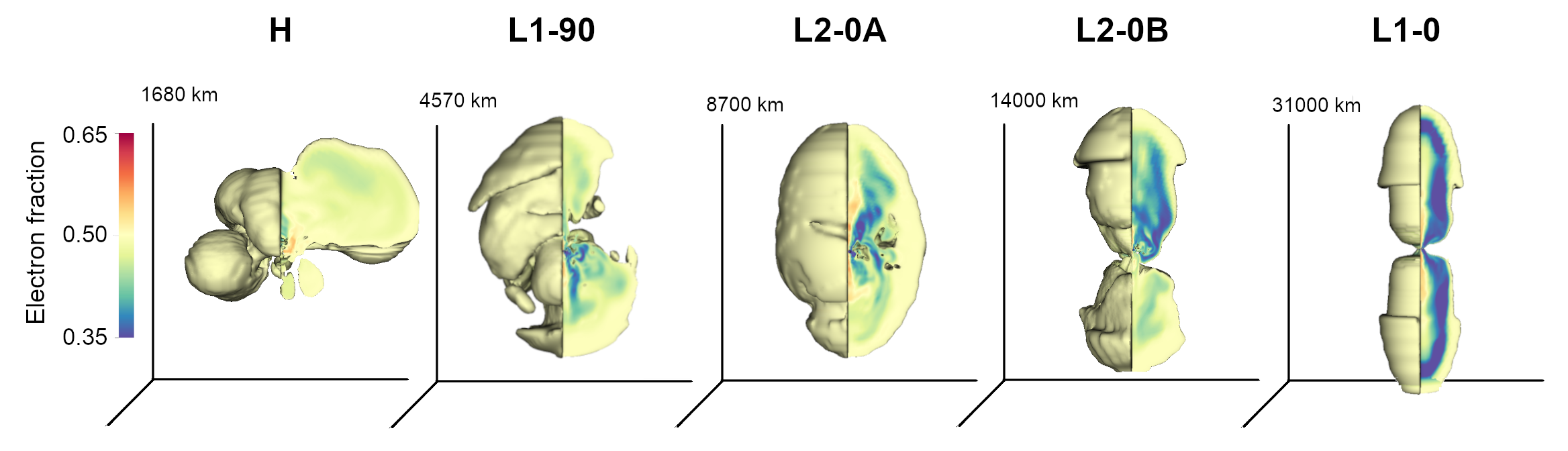}%
\end{center}
\caption{Electron fraction of the ejecta from the different models at the end of the simulation. The models are roughly ordered from least (left) to most (right) jet-like. This coincides with the neutron-richness of the model.} 
\label{fig:ye_models}
\end{figure*}

\subsection{Nucleosynthesis calculations}
We use an updated version of the nuclear reaction network \mbox{\textsc{WinNet}} (\citealt{Winteler2012}, \citealt{winnet}) with the same setup as in \citet{Reichert2023}. This network considers $\sim 6500$ nuclei up to $Z=111$. We include fission fragment distributions and reactions from \citet{Panov2001, Panov2005,Panov2010}. Spontaneous fission half-lives are included as in \citet{petermann2012}. Additionally, we include (anti-)neutrino cross sections on nucleons as described in \citet[][]{Burrows2006} with weak magnetism and recoil corrections of \citet[][]{Horowitz2002}. We consider reaction rates from the Reaclib library \citep{Cyburt2010}. Theoretical $\beta^+$, $\beta^-$, electron capture, and positron capture rates at stellar conditions are taken from \citet{Langanke2001b}. These rates are exchanged for experimental rates included in the Reaclib library at $T=0.01\,\mathrm{GK}$. All calculations are performed for $1\, \mathrm{Gyr}$ to ensure that unstable nuclei have decayed to stability. In a post-processing stage, we sample the neutrino-MHD simulations employing Lagrangian markers, hereafter called \emph{tracers}. All tracers are placed at the end of the simulation and integrated backwards in time (see e.g., \citealt{Sieverding2023} for an overview of the method uncertainties). For all models, we group the tracers that sample the ejecta in electron fraction and entropy (two-dimensional) bins. The neutron-richness, or electron fraction, is the main parameter for the synthesis of heavy elements. As demonstrated in \citet{Reichert2023}, it is possible to calculate only a subset of trajectories within these groups in order to get the total yields. We therefore calculate only $25$ tracers per two-dimensional bin as in \citet{Reichert2023}.

\section{Results}
In the following, we look at the nucleosynthesis of all investigated models and discuss the impact of different initial magnetic field topologies. This supplements the previous work of \citet[][]{Reichert2023} who investigated the impact of different magnetic field strength and various axially symmetric magnetic topologies of 3D models of the very same progenitor. Afterwards, we discuss the impact of nuclear physics input on our results.

\subsection{Impact of different magnetic field configurations on the yields}
\label{sct:yield}
First, we want to focus on how different magnetic field configurations can have an impact on the nucleosynthesis. 
When looking at the integrated abundances in Fig.~\ref{fig:nucleosynthesis_frdm}, one immediately notices the difference between model H and any other model including magnetic fields, the latter producing heavy elements beyond the 1st r-process peak. More specifically, we can notice the difference between the aligned dipole model L1-0 and model H in Fig.~\ref{fig:nucleosynthesis_frdm}, as the two extreme cases bounding the influence of initially poloidal fields in our models. While the purely dipolar model produces heavy elements up to the 3rd r-process peak, the model without magnetic fields does not produce any yield beyond the 1st r-process peak. The rest of the magnetized models display abundance patterns in between of the two bounding cases. The production of heavy elements when strong dipolar fields are included in the initial model is a robust feature consistently found also in previous works (\citealt{Nishimura2006,Nishimura2015,Nishimura2017,Moesta2018,Halevi2018,Reichert2021a,Reichert2023}). Model H shows a very narrow distribution of the neutron-richness around a symmetric value of $Y_e=0.5$ (Fig.~\ref{fig:ye_time_hist}). The most neutron-rich matter in this model has still a value of $Y_e \sim 0.4$, not enough to significantly synthesize elements with mass numbers beyond $A>100$. Slightly heavier nuclei until $A\sim 140$ are reached by all models that include a magnetic field. Hereby, the yields of the models that involve a quadrupolar setup (models L2-0A and L2-0B) and the model with 90 degree tilted dipole show very similar yields. This similarity is due to waiting points that have to be overcome such as the closed neutron shell at $N=50$. In general, a more jet-like structure (see Fig.~\ref{fig:ye_models}) leads to more neutron-rich conditions and therefore more favorable to produce heavier nuclei. 
\begin{figure}
\begin{center}
\includegraphics[width=1.0\linewidth]{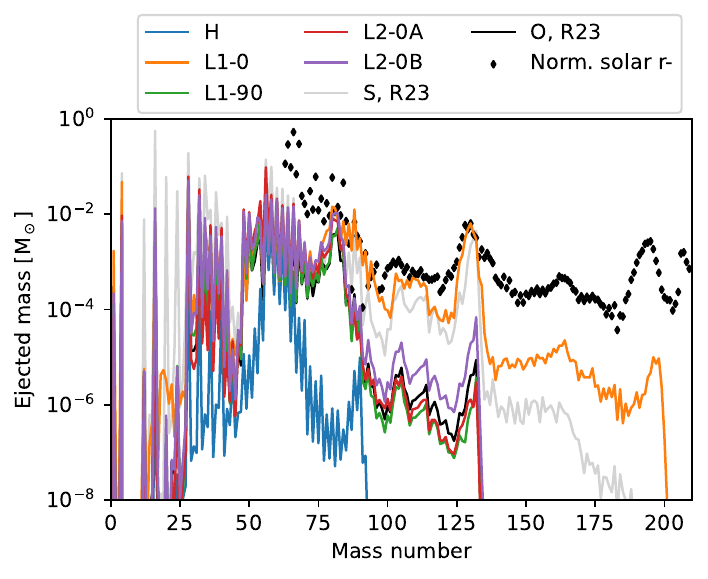}%
\end{center}
\caption{Ejecta composition at the end of the simulations using $(n,\gamma)$ and $(\gamma,n)$ rates of the JINA Reaclib. Only the model with the global aligned dipole is able to eject heavy elements with $A>150$. For comparison, we also show model O and S that was presented in \citet[][labeled in the figure with O, R23 and S, R23, respectively]{Reichert2023}. } 
\label{fig:nucleosynthesis_frdm}
\end{figure}
The large-scale dipole model L1-0 is the only one that achieves electron fractions below $Y_e<0.25$. These conditions are also only reached very shortly after bounce ($t_\mathrm{pb}\sim 0.05\,\mathrm{s}$). At later times, most of the ejected matter has electron fractions of $\sim 0.38$. Hence, in agreement with our previous results, the yields of r-process elements are produced by the so-called \emph{prompt} mechanism \citep{Reichert2023}.

An interesting comparison is also given by the difference between the yields of model L1-0 and model S of \citet[][grey line in Fig.~\ref{fig:nucleosynthesis_frdm}]{Reichert2023}. While the magnetic poloidal topology and the progenitor are the same, there is a difference in the magnetic field setup. Model S starts with a toroidal field component in equipartition with the poloidal magnetic component, while model L1-0 starts without this component \citep{Obergaulinger2020,Bugli2021}. As a result, there are changes in the neutron-richness to a level that is visible in the yields (compare grey and orange line in Fig.~\ref{fig:nucleosynthesis_frdm}) because the electron fraction is close to the critical value that enables an r-process beyond the second peak. The variation in the abundances between the two models may be slightly enhanced by the different resolution of the tracer particles (an average of $1.08\times 10^{-6}\,\mathrm{M}_\odot$ versus $1.59\times 10^{-7}\,\mathrm{M}_\odot$ per tracer particle for model S and L1-0, respectively). However, we expect the main difference coming from the magnetic field as shown also in the MHD simulations (compare Fig.~\ref{fig:ye_models} with Fig.~1 of \citealt{Reichert2023}). Our calculations are extremely sensitive to small changes because the nucleosynthetic flow is right at the threshold of being able to synthesis a full r-process up to the third r-process peak. 

For comparison, Fig.~\ref{fig:nucleosynthesis_frdm} also shows the yields of model O presented in \citet[][]{Reichert2023}. This model hosts a magnetic field that is consistent with the underlying progenitor model. The magnetic field of this model is a combination of many local dipolar fields with non-magnetized layers in between. This could explain the similarity of its nucleosynthetic yields to those produced by the models with magnetic topology other than the large-scale aligned dipole.

\begin{figure}
\begin{center}
\includegraphics[width=1.0\linewidth]{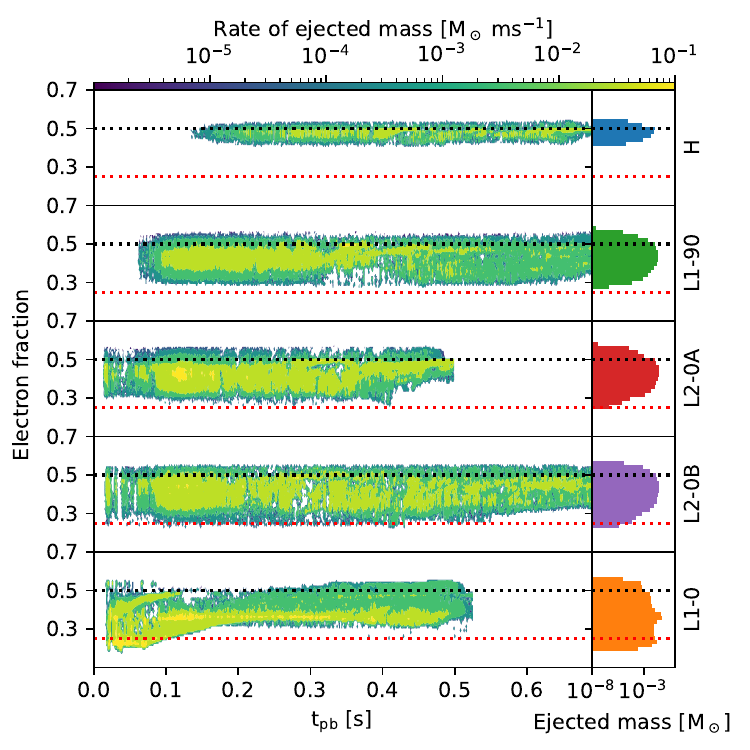}%
\end{center}
\caption{Left panels show the rate of ejected mass (according to the upper colour scale) as a function of time from bounce when the tracer is ejected. The results are  binned according to the electron fraction at the time when the tracers reach a temperature of $7\, \mathrm{GK}$ (which may happen anytime since the beginning of the tracer trajectories, not necessarily at the time after bounce at which they are ejected). Tracers that never reach $7\, \mathrm{GK}$ along their evolution are not plotted here. The right panels show the integrated yields binned in electron fraction at $7\,\mathrm{GK}$. Row wise, each panel shows a different model that is indicated with the label on the right side of the plot. The horizontal black dotted lines indicate a value of $Y_e=0.5$. The red dotted lines denote $Y_e=0.25$, a value that roughly indicates the conditions for the synthesis of heavy elements with $A>150$. Only the aligned dipole model L1-0 reaches sufficiently low electron fractions to synthesize heavier elements.} 
\label{fig:ye_time_hist}
\end{figure}

\begin{table}
\centering
\begin{tabular}{l l l l}
\hline
\hline
\textbf{Model} & \textbf{$^{44}$Ti} & \textbf{$^{56}$Ni} & \textbf{$^{60}$Fe}\\ 
\hline 
H     &  $2.17 \cdot 10^{-6}$ & $6.12\cdot 10^{-3}$ & $2.57\cdot 10^{-5}$\\ 
L1-90 &  $1.39 \cdot 10^{-5}$ & $3.45\cdot 10^{-2}$ & $4.08\cdot 10^{-4}$\\ 
L2-0A &  $1.92 \cdot 10^{-5}$ & $8.28\cdot 10^{-2}$ & $4.44\cdot 10^{-3}$\\ 
L2-0B &  $1.21 \cdot 10^{-5}$ & $1.86\cdot 10^{-2}$ & $4.18\cdot 10^{-3}$\\ 
L1-0  &  $3.14 \cdot 10^{-6}$ & $9.07\cdot 10^{-3}$ & $4.97\cdot 10^{-4}$\\ 
\hline        
\end{tabular}
\caption{Ejecta yields in solar masses of unstable isotopes taken at one tenth of their half life. The values given here are lower limits, as we can only track their growth up to the end of the simulation. Note that there will also be a contribution to $^{60}$Fe from the progenitor that is produced during stellar evolution and not included in the values given here \citep[see, e.g.,][for a discussion]{Diehl2021,Reichert2023}{}{}}
\label{tab:unstable_ejecta}
\end{table}
Besides the total yield, we also calculated the ejecta yields of the unstable nuclei $^{44}$Ti, $^{56}$Ni, and $^{60}$Fe at the end of the simulations (Tab.~\ref{tab:unstable_ejecta}). These values have to be taken with care as they only take into account the ejecta that was present at the end of the simulation (Tab.~\ref{tab:sim_properties}). The order of magnitude of the ejected $^{44}$Ti is about $10^{-5}\,\mathrm{M_\odot}$, similar to the models calculated in \citet[][]{Reichert2023}. There, they estimated that the amount of $^{44}$Ti may still grow by possibly one order of magnitude. Therefore, all the values obtained here could still be consistent to the ones of observed CC-SNe such as SN1987A \citep[$5.5 \cdot 10^{-5}\, \mathrm{M_\odot}$,][]{Seitenzahl2014} or Cas A \citep[$1.3\cdot 10^{-4}\, \mathrm{M_\odot}$,][]{WangLi2016}. At first glance, it seems interesting that the most extreme models (model without magnetic field (H) and the one hosting a large scale dipole (L1-0)) produce similar amount of $^{44}$Ti. $^{44}$Ti needs close to symmetric matter (i.e., $Y_e\sim 0.5$) to be synthesized and only a small amount of the ejecta in L1-0 reaches these condition as it is dominated by neutron-rich matter. The other extreme model, H, ejects mainly symmetric matter however the total unbound mass is smaller at the end of the simulation, see Tab.~\ref{tab:sim_properties}.

\begin{figure}
\begin{center}
\includegraphics[width=1.0\linewidth]{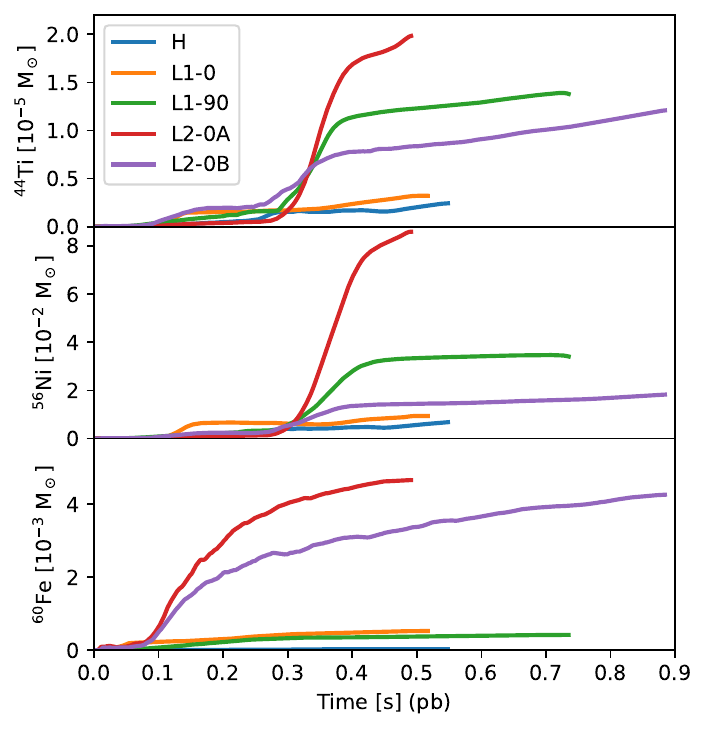}%
\end{center}
\caption{Ejected mass of unstable nuclei. From top to bottom, the panels show the ejected mass of $^{44}\mathrm{Ti}$, $^{56}\mathrm{Ni}$, and $^{60}\mathrm{Fe}$, respectively.}
\label{fig:ni_vs_time}
\end{figure}
The amount of $^{56}$Ni falls within the range $\sim 6\cdot 10^{-3}\,\mathrm{M_\odot} - 8\cdot 10^{-2}\,\mathrm{M_\odot}$ which is consistent with CC-SNe observations \citep[c.f., $7 \cdot 10^{-2}\, \mathrm{M_\odot}$,][for SN1987A]{Seitenzahl2014}, but also most likely will continue to grow with increasing simulation time (Fig.~\ref{fig:ni_vs_time}). As in the case of the synthesised $^{44}$Ti, also for $^{56}$Ni we find a larger amount in models L1-90, L2-0A, and L2-0B (Fig.~\ref{fig:ni_vs_time}). The fact that this trend is analogous to the one found for $^{44}\mathrm{Ti}$ can be explained by the similar formation conditions of $^{44}\mathrm{Ti}$ and $^{56}\mathrm{Ni}$ (both require $Y_e \sim 0.5$). The amount of ejected $^{56}$Ni is especially large in model L2-0A, hosting a quadrupolar magnetic field aligned with the rotational axis. This model explodes more spherical than the models L2-0B and L1-0 (Fig.~\ref{fig:ye_models}) and most of the (hot, i.e., $T>5\,\mathrm{GK}$) ejected matter is symmetric. This can be explained by the shock being energetic enough to unbound the matter in the equatorial region before the electron fraction gets majorly altered from its initial symmetric value stemming from the progenitor. At the same time, this model is ejecting a larger amount of matter than model H and L1-90 (Tab.~\ref{tab:sim_properties}). Notably, this leads also to a around twice as large ejection of $^{56}$Ni compared to the other quadrupole model with a more approximate gravitational treatment (L2-0B).

In comparison to the results from model Sof \cite{Reichert2023} at $t_\mathrm{pb}\approx 0.4\,$s, model L1-0 produces about one order of magnitude less $^{56}$Ni suggesting that the existence of a toroidal component of the magnetic field in the stellar progenitor has also a significant effect on the supernova lightcurve of the event that is powered by the decay of $^{56}$Ni. Whether this is a robust feature in the absence of toroidal fields or a statistical feature of the models has to be proven with additional simulations. Within existing literature, \citet[][]{Nishimura2017} reported a low amount of ejected $^{56}$Ni (see their Fig.~5), also not assuming an initial toroidal field. This is consistent with our findings. 

The overabundance of $^{60}$Fe in models with magnetic fields (in comparison to the H model) is remarkable. Quadrupolar magnetic fields in the progenitor may boost the production of this element by more than two orders of magnitude with respect to the purely hydrodynamic model. Comparing the yields of model L1-0 with the ones of model S of \cite{Reichert2023}, the latter produces about 10 times more $^{60}$Fe. This is a direct consequence of having more ejected mass with electron fractions of $\sim 0.43$ in model S ($\sim5.1\times 10^{-3}\, \mathrm{M_\odot}$ and $\sim1.8\times 10^{-2}\, \mathrm{M_\odot}$ for ejecta with \mbox{$0.42<Y_e<0.44$} at $7\, \mathrm{GK}$ and \mbox{$t_\mathrm{pb}=0.525\,\mathrm{s}$} for model L1-0 and S, respectively) as $^{60}$Fe is dominantly produced in these conditions. Thus, we consider initial toroidal fields as very important for the production of $^{60}$Fe. We note that the amount of $^{60}$Fe discussed here does not include the contribution that is synthesized during stellar evolution \citep[e.g.,][]{Diehl2021}{}{}. This could be the dominant contribution to $^{60}$Fe for all of our models (see \citealt{Reichert2023} for a more detailed discussion).

In total, the effect of the initial magnetic field topology is as important as the field strength itself (c.f., \citealt{Reichert2023}). For the large field strength assumed here, we at least synthesize elements up to the second r-process peak (A$\sim 130$).

\subsection{The impact of nuclear physic uncertainties}
\label{sct:nuc_phys_unc}
Besides changes in the magnetic field topology, there is also an uncertainty coming from the nuclear physics.
In the following, we investigate the variance of the results coming from these uncertainties. This spans the impact of unknown nuclear masses, $\beta^-$-decay half-lives and $\beta^-$-delayed neutron emission probabilities, neutrino reactions, and nuclear energy generation that can feedback on the temperature. We do not investigate the impact of fission reactions and fragments because none of our models is sufficiently neutron-rich for fission to play a significant role, and we therefore ignore the related uncertainties. Within this section, we will change individual nuclear inputs and compare the difference to the yields that we presented in the last section. Since model L2-0B is similar to the yields of model L1-90 and L2-0A and includes a more approximate gravitational potential, we exclude this model from the following analysis.

\subsubsection{The impact of nuclear masses}
First, we will discuss the impact of using different mass models for calculating $(n,\gamma)$ and $(\gamma,n)$ reaction rates. Following \citet{Martin2016}, we use same masses and calculated rates. The underlying nuclear masses were originally from \citet{Erler2012} and include six different energy density functionals\footnote{Access via \url{http://massexplorer.frib.msu.edu/}}, SKM$^*$ \citep{Bartel1982}, SkP \citep{Dobaczewski1984}, SLy4 \citep{Chabanat1998}, SV-min \citep{Kluepfel2009}, UNEDF0 \citep{Kortelainen2010}, and UNEDF1 \citep{Kortelainen2012}. The nuclear masses are critical for $r$-process calculations as a $(n,\gamma)-(\gamma,n)$ equilibrium  is established and the path is determined by the neutron separation energies \citep{Kratz1993}. We only consider the nuclear mass variations in the calculation of $(n,\gamma)$ and $(\gamma,n)$ reactions as this has the largest impact.

For our conditions, the impact of nuclear masses on the synthesis of heavy elements is therefore fairly small. As illustrated in Fig.~\ref{fig:l1b12model_chart} the nucleosynthetic flow of the most neutron-rich model, L1-0, moves  along nuclei with experimentally known masses.  Most of the nucleosynthetic flow is located in a region of known nuclear masses \citep[][region between the red lines]{Tuli2011}. However, matter is located outside this region for the second and third r-process peaks at the neutron magic numbers $82$ and $126$ (indicated as black dashed lines). Therefore, the impact of the underlying DFT masses is most visible around the peaks (Fig.~\ref{fig:ngamma}), where the largest error can be seen at the third r-process peak. This is in qualitative agreement with the results presented in \citet[][]{Martin2016}. A couple of considerations have to be taken into account here to compare our results to the ones of \cite{Martin2016}. First, they used a single trajectory from \citet[][]{Winteler2012} while our Fig.~\ref{fig:ngamma} displays the cumulative results of $\sim 2000$ trajectories. Second, the simulation of \citet[][]{Winteler2012} was run for less than $40$\,ms post-bounce, while our L1-0 model has been run for $\sim 0.5$\,s post-bounce. In contrast to \citet[][]{Winteler2012}, the simulations here include improved neutrino transport and this leads to less neutron-rich conditions. Therefore, the new estimates presented here represent a major improvement.
\begin{figure}
\begin{center}
\includegraphics[width=1.0\linewidth]{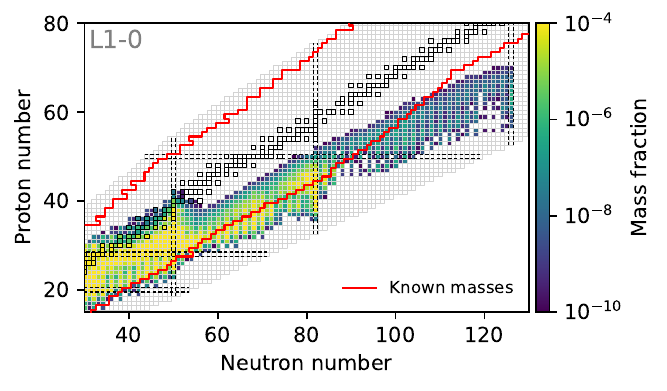}%
\end{center}
\caption{Mass fractions at the end of the simulation of model L1-0 ($t_\mathrm{f,pb}=0.525\,\mathrm{s}$) in the nuclear chart. Shown is the weighted average of all calculated tracers. Experimentally measured nuclear masses lie in the region between the two red lines \citep[][]{Tuli2011}. Magic numbers are indicated as dashed lines.} 
\label{fig:l1b12model_chart}
\end{figure}

\begin{figure}
\begin{center}
\includegraphics[width=1.0\linewidth]{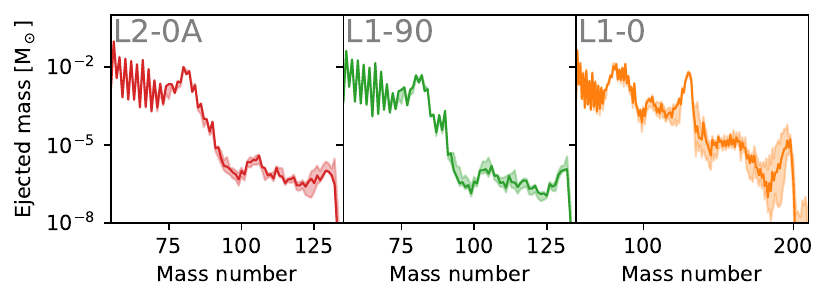}%
\end{center}
\caption{Ejected mass using seven different sets of nuclear masses for the calculation of $(n,\gamma)$ and $(\gamma,n)$ reaction rates (see text). Shaded regions indicate the maximum and minimum yield. Solid lines indicate the median of all yields. The different panels show models L2-0A, L1-90, and L1-0. Note that the range of the mass number differs in the right panel compared to the other ones.} 
\label{fig:ngamma}
\end{figure}
In contrast to \citet[][]{Winteler2012}, the simulations here host much less neutron-rich conditions due to the improved neutrino transport in the underlying neutrino-MHD simulations. So we consider our estimates here a major improvement compared to the results presented in the aforementioned studies.

\subsubsection{The impact of $\beta^{-}$-decay half-lives and $\beta^{-}$-delayed neutron emission probabilities}
\begin{figure}
\begin{center}
\includegraphics[width=1.0\linewidth]{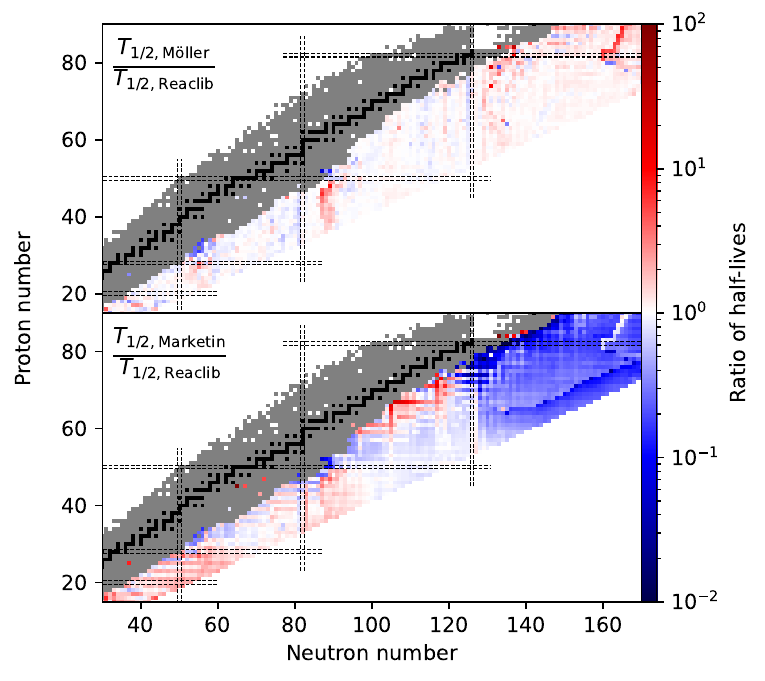}%
\end{center}
\caption{Ratios of $\beta^-$-decay half-lives for \citet[][]{Moeller2019} and \citet[][]{Marketin2016} with respect to the half-lives included in the JINA Reaclib that mostly originate from \citet{Moeller2003}. Blue rectangles indicate faster decays compared to the ones included in the JINA Reaclib, red rectangles indicate slower decays. Experimentally determined $\beta^{-}$-decays are shown as grey rectangles, stable nuclei as black rectangles. The magic numbers are plotted as horizontal and vertical dashed lines.} 
\label{fig:betahalflives}
\end{figure}
\begin{figure}
\begin{center}
\includegraphics[width=1.0\linewidth]{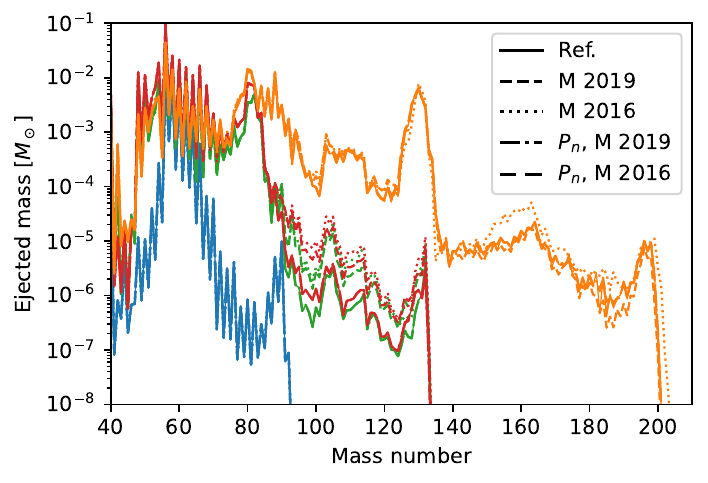}%
\end{center}
\caption{Ejected masses for different $\beta^{-}$-decays indicated by the line style. The different colours indicate the models, H (blue), L2-0A (red), L1-90 (green), L1-0 (orange). The reference (Ref., straight lines) includes $\beta^{-}$-decays from the JINA Reaclib \citep[][]{Cyburt2010}. Results with $\beta^{-}$-decays from \citet[][]{Moeller2019} are shown as short dashed lines, with decays from \citet[][]{Marketin2016} as dotted lines. Additionally, results for $\beta^-$-decays with half-lives as in the JINA Reaclib, but $\beta^-$-delayed neutron emission probabilities of \citet[][]{Moeller2019} and \citet[][]{Marketin2016} are shown as dashed-dotted and long dashed lines, respectively. These lines are so close to the reference (Ref.) that they are not visible. } 
\label{fig:beta_yields}
\end{figure}

Beta decays set the speed of the r-process determining how fast and how much matter is moved towards heavier nuclei \citep[see, e.g.,][]{Hosmer2010,Madurga2012,Eichler2015,Lund2023}. Here, we replaced the $\beta^-$-decays of the Reaclib \citep[originating from][]{Moeller2003} with the ones of \citet[][]{Marketin2016} and \citet[][]{Moeller2019}. Hereby, we ensured that we only replace decays that are not experimentally known. The different half-lives are shown in Fig.~\ref{fig:betahalflives}. In addition, the channels of the decay can differ, as a $\beta^{-}$-decay may be ensued by the release of several neutrons. This gets increasingly important close to the neutron-drip line. All three sets of $\beta^{-}$-decays that we use here predict a different amount of $\beta^{-}$-delayed neutrons. Within the JINA-Reaclib, at most $2$ neutrons can be emitted, within \citet[][]{Marketin2016} a maximum of $5$, and within \citet[][]{Moeller2019} a maximum of $10$. We note that not only the maximum amount of neutrons differs, but more strikingly the average amount of ejected neutrons per decay.

Figure~\ref{fig:beta_yields} shows the impact of the different decay rates. While models H and L1-0 show only slight deviations for the different rates, the remaining magnetized models L1-90 and L2-0A show an increased yield for elements with masses $90\lesssim A \lesssim 130$ (of less than one order of magnitude).

To disentangle the contribution of different half-lives from the contribution of different $\beta^{-}$-delayed neutron emission probabilities, we calculated all models with the half-lives from the JINA Reaclib, but using the probabilities for $n$ $\beta^{-}$-delayed neutrons $P_n$ of \citet[][]{Marketin2016} and \citet[][]{Moeller2019}. It turns out that the impact of $\beta^{-}$-delayed neutrons is negligible for our models (Fig.~\ref{fig:beta_yields}).

The difference of the abundances is mainly due to the shorter half-lives (blue rectangles in Fig.~\ref{fig:betahalflives}) in the region of $51 < N < 59$ and $29 < Z < 35$. It was already shown in, e.g., \citet{Eichler2015} that faster $\beta^{-}$-decays lead to a  faster flux and therefore can lead to a larger amount of heavier elements. Furthermore, $\beta^{-}$-decays are very important close to the neutron shell closures (e.g., $N\sim 50, 82$, or $126$). The half-lives in these regions are similar for \citet[][]{Marketin2016} and \citet[][]{Moeller2019}, but around one order of magnitude smaller than the one of \citet[][]{Moeller2003} that are included in the JINA Reaclib. Therefore, the effect on the abundances is similar for both rate tables, \citet[][]{Marketin2016} and \citet[][]{Moeller2019}. The uncertainty of these decays may greatly reduce when new half-lives in this region become available (e.g., \citealt{Madurga2012,Hall2021,Phong2022,Yokoyama2023}).

\subsubsection{The impact of neutrino reactions}
Neutrinos can have a major impact on the nucleosynthetic yields, as charged-current reactions can modify the neutron-richness of the matter. We calculated the abundances of all models, saved L2-0B, for four different treatments of neutrinos. This includes charged-current neutrino reactions only on nucleons, no neutrino reactions, charged-current reactions on nucleons and heavier nuclei as in \citet[][]{Froehlich2006}, and the same reactions with rate tables from \citet[][]{Sieverding2018}. We stress that when changing or turning off the neutrino reactions, we only do this for the tracers after they reach a temperature of $7\, \mathrm{GK}$. The nucleosynthesis calculation nevertheless starts with an initial electron fraction that is taken from the simulation at $7\, \mathrm{GK}$ and that includes the neutrino treatment that was used in the neutrino-MHD simulation (i.e., using the M1 transport scheme with a simplified treatment of neutrino charged-current reactions on heavy nuclei as described in \citealt{Cernohorsky1992} or \citealt{Bruenn1985}). 

The impact of the different neutrino reaction rates can be seen in Fig.~\ref{fig:neutrinos}. Model H shows a sensitivity on neutrino reactions on heavy nuclei in a region around mass numbers of $80 \lesssim A\lesssim 100$. Hereby, the rate tables of \citet[][]{Froehlich2006} and \citet[][]{Sieverding2018} lead to similar results. This shows that the additional channels of the rates of \citet[][]{Sieverding2018} which include charged-current reactions with an additional ejection of neutrons, protons, or alpha particles are not too relevant for our conditions. While model L2-0A is rather insensitive to the changes in neutrino reactions, model L1-90 also shows an increase of the abundance of elements with masses around $A\sim 100$. Interestingly, model L1-0 only shows a minor sensitivity to neutrino reactions. In fact, there is only a slight difference visible between taking neutrino reactions into account and not considering any neutrino reactions (dashed line in Fig.~\ref{fig:neutrinos}). Within the simulation, the neutrino properties are such that they usually increase the electron fraction. Therefore, without neutrino reactions in the network, we get slightly more neutron-rich material which manifests in an increased amount of ejected third r-process peak material. The effect is nevertheless small as most neutron-rich matter is ejected early on and fast. So fast, that the neutrinos only play a minor role as the electron fraction was already set at higher temperatures ($T>7\,$GK, i.e., before we start the nucleosynthesis calculation). 
\begin{figure}
\begin{center}
\includegraphics[width=1.0\linewidth]{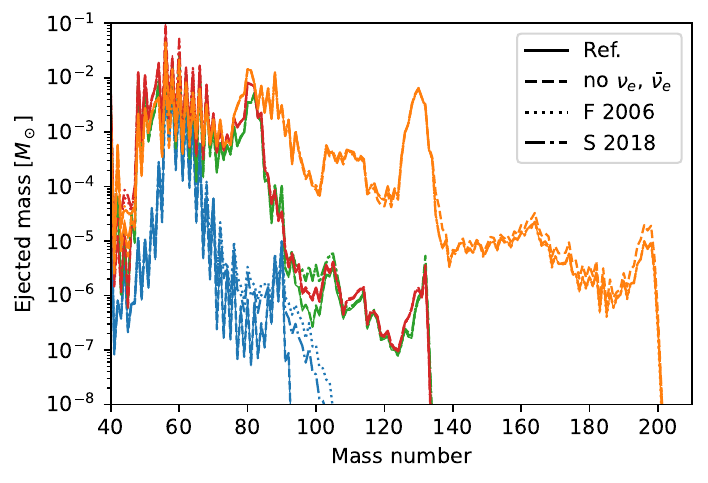}%
\end{center}
\caption{Ejected masses for different neutrino treatments indicated by the line style. The different colours indicate the models, H (blue), L2-0A (red), L1-90 (green), L1-0 (orange). The reference (Ref., solid lines) includes neutrino reactions on nucleons only, no neutrino reactions are shown as dashed lines, dotted lines indicate neutrino reactions on nucleons and charged-current reactions on heavier nuclei as in \citet{Froehlich2006}, and dashed dotted lines show the result with neutrino reactions on nucleons and charged-current reactions on heavier nuclei as in \citet{Sieverding2018}. For model L1-0 and L2-0A, the reference run lies on top of the calculations including neutrino reactions on heavier nuclei.}  
\label{fig:neutrinos}
\end{figure}

\subsubsection{The impact of nuclear energy}
Energy released by nuclear reactions has a feedback on the temperature of the surrounding matter. While the tracers track the temperature within the simulation, we can also post-process the temperature. For this, we use an approach as described in, e.g., \citet[][]{Mueller1986}, \citet{Lippuner2017} or \citet[][]{winnet}. From the first law of thermodynamics, one can derive
\begin{equation}
\label{eq:entropy_dgl}
    \Delta{S} = -\frac{ 1}{k_\mathrm{B}T}\sum _i (\mu _i + Z_i \mu _e) \Delta Y_i - \dot{q}.
\end{equation}
Hereby, $S$ is the entropy, $T$ the temperature, $\mu_i$ the chemical potential of nucleus $i$, $Z_i$ its proton number, $\mu_e$ the chemical potential of the electron, and $\Delta Y$ the abundance change within the time step. The energy added and lost by the system is included in $\dot{q}$. Here we assume that energy is lost in form of neutrinos by using the tabulation of \citet[][]{Langanke2001b} for theoretical weak rates, a table of average neutrino energies from $\beta^{-}$-decays from experimentally known decays that we extract from the ENSDF database \citep[][]{Brown2018}. If a decay is not included in either of the tables, we assume that $40\%$ of the Q-value of the reaction is released in form of neutrinos \citep[as e.g., suggested by the tables of ][]{Marketin2016}. Of minor importance, we include the energy loss of thermal neutrinos with the parametrization of \citet[][]{Itoh1996}. Additionally, we include that energy can enter the system via neutrino reactions on nucleons where we assume that the average energy of the absorbed neutrino is added.
Self-heating as described here has the advantage that the temperature includes feedback from the nuclear reactions. Furthermore, it can cure uncertainties within the tracer particle integration by connecting the density to a physical meaningful temperature. On the other hand, we are neglecting non-adiabatic effects that could come from shocks.

For all models, we observe a negligible effect from nuclear heating on the abundance pattern (Fig.~\ref{fig:heating}). 
This is reassuring as it indicates that our tracer particle integration is reliable. Furthermore, it shows that the dynamics is not majorly impacted by nuclear reactions. 
The largest, even though still small, impact can be seen in model L1-0. This is also expected, as heating by $\beta^{-}$-decays happens for the involved neutron-richness in model L1-0 at small timescales that are not negligible compared to the simulation time. Such effects could only be included in the MHD simulations by the incorporation of a large network with thousands of nuclei, given the neutron-richness involved. This is nowadays not feasible and at most reaction networks with at most a couple of hundreds nuclei have been included into simulations \citep[e.g.,][]{Harris2017,Navo_2022arXiv221011848,Sandoval2021}.

\begin{figure}
\begin{center}
\includegraphics[width=1.0\linewidth]{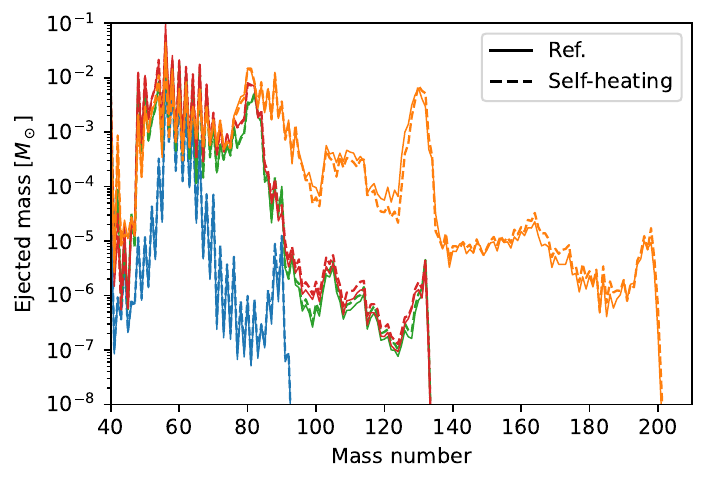}%
\end{center}
 \caption{Impact of self-heating. Shown are the ejected masses of model H (blue), L1-90 (green), L2-0A (red), and L1-0 (orange) as a function of mass number. The reference calculations without self-heating taken into account are the solid lines, identical to those in Fig.~\ref{fig:nucleosynthesis_frdm}. Self-heating marginally effects model L1-0 while other models are rather insensitive.} 
\label{fig:heating}
\end{figure}

\section{Implications from observations}
\label{sct:observations}
When comparing theoretical yields with observations of abundances of elements in stellar atmospheres, one has to be aware of several factors that may influence this comparison, i.e., 
\begin{enumerate}
\item \emph{Progenitor and explosion models}. The astrophysical uncertainties that arise through physical approximations employed in the modeling of progenitors and in magnetohydrodynamic simulations. Among these astrophysical uncertainties we single out the magnetic field strength \citep[as investigated in e.g.,][]{Nishimura2017,Reichert2023}, the orientation of the magnetic field relative to the rotational axis \citep[as investigated here and in][]{Halevi2018}, and the magnetic field topology. Furthermore, independent of uncertainties, different properties of the progenitor such as masses or metallicities may lead to different nucleosynthetic outcomes.
\item \emph{Nuclear physics}. The nuclear physics input have been discussed in the previous section. Compared to the astrophysical uncertainties they are comparably small as our models do not reach very neutron-rich conditions and the nucleosynthesis path therefore moves along nuclei with experimentally known masses.
\item \emph{Observations of stellar atmospheres} The uncertainties and constraints that arise when observing stellar atmospheres. The observation of some elements can have large uncertainties due to assumptions for the stellar atmosphere, such as local thermodynamic equilibrium (LTE) or restricting them to spherical symmetry \citep[e.g.,][]{Castelli2003}. Additionally, the atomic physics properties of some transitions may have not been experimentally precisely measured, further increasing the error of the abundances \citep[e.g.,][]{Gray2005}. Even worse, some elements are not observable due to unfavourable atomic properties and could only be detected in a wavelength range that is disturbed by Earth's atmosphere, making it necessary to observe them with telescopes that are located in space. This vastly decreases the amount of detections and, therefore, decreases the amount of available data (see top panel and background colors in the lower panel of Fig.~\ref{fig:abundances_obs}). 
\end{enumerate}  

After summarising these three sources of uncertainties and variations in the calculated and observed abundances, we want to discuss the potential of MR-SN to account for the variability found in observed abundances. Elemental r-process abundances of the oldest stars show two characteristic trends represented by the two stars in Fig.~\ref{fig:abundances_obs}.  There are stars with high and low enrichment of heavy elements from second to third peak relative to first peak (Sr-Y-Zr, \citealt{Aoki2005, Roederer2010, Hansen2014, Qian2001a,Qian2007}). Those with high abundances for heavy r-process elements present also a robust pattern, namely the relative scale of abundances are the same in all those stars and in the solar system \citep[e.g.,][]{Sneden2003, Sneden2008, Roederer2018}. However, for lighter heavy elements between Sr and Ag the pattern is less robust. This together with the observation of stars with low enrichment of heavy r-process indicates that there may be an additional contribution to those elements from another astrophysical site(s). Such an addition contribution was discussed by \citet{Qian2001a} and \citet{Travaglio2004} who called it LEPP (lighter element primary process). There has been several suggestions about an extra astrophysical site to produce elements between Sr and Ag in addition to the r- and s-process \cite{Montes2007}: fast rotating stars \citep[e.g.,][]{Frischknecht2012,Frischknecht2016}, neutrino-driven supernova \citep[e.g.,][]{Arcones2011,Hansen2014,Bliss2018,Witt2021,Sieverding2023b,Wang2023}, neutrino-driven winds after neutron star mergers \citep[e.g.,][]{Martin2015,Fujibayashi2017}.

Our results show that MR-SNe could also contribute to the production of lighter heavy elements and explain the variations found in the observed abundances. Moreover, they  may explain the stars with low abundance of heavy r-process elements. When comparing the yields to strong r-process enhanced stars like HD 222925 \citep[][black squares in Fig.~\ref{fig:abundances_obs}]{Roederer2022} it becomes clear that even in the very optimistic case of an aligned dipole we can not reproduce the abundances for elements heavier than Barium ($Z=56$). Also in the case of an only slightly r-process enriched star, HD 122563 \citep[][magenta circles in Fig.~\ref{fig:abundances_obs}]{Honda2007}, heavy elements are under-produced. This underproduction of heavy r-process elements compared to lighter ones (between Sr and Ag) is a common problem in MR-SNe \citep[see also e.g.,][for a discussion]{Ekanger2023}. This may be enhanced when considering longer simulation times because at late times the production of heavy elements is negligible while lighter heavy elements are still synthesised (see the continuous ejection of material with $Y_e\lesssim 0.45$ in Fig.~\ref{fig:ye_time_hist} and the discussion in \citealt{Reichert2023}).
If stars like HD 122563 are indeed carrying the nucleosynthetic fingerprints of MR-SNe, there should be a mechanism that prevents a further ejection of neutron-rich matter at later times (see Fig.~\ref{fig:ye_time_hist}), i.e. fall-back or earlier black hole formation. 
We also want to note that the potential for a postponed formation of a collapsar could result in an additional ejection of r-process material. The question of whether this phenomenon would bring the ejected material closer to resembling the pattern observed in robust r-process stars like CS22897-008, or exacerbate the issue of overproduction of lighter heavy elements, remains a subject of ongoing research (see e.g., \citealt{Siegel2019} and \citealt{Miller2020}).

\begin{figure}
\begin{center}
\includegraphics[width=1.0\linewidth]{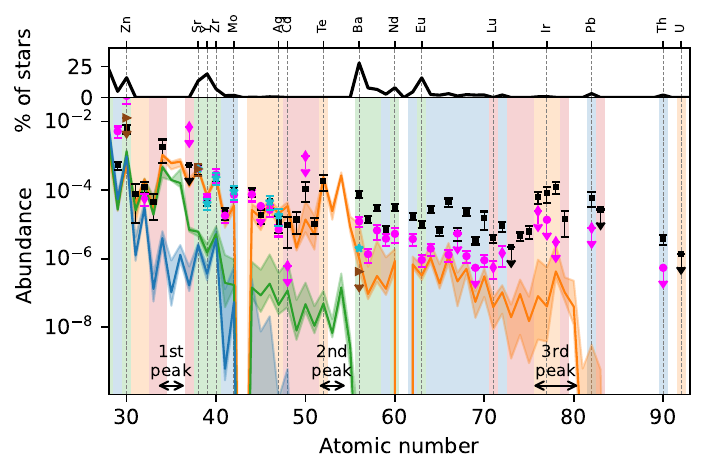}%
\end{center}
\caption{Upper panel: Percentage of observations per element according to the SAGA database, adding the detections of the star HD 222925. Dashed vertical lines show specific elements as indicated at the top of the plot. Lower panel: Median abundances of all previous calculations of the models H (blue line), L1-90 (green line), and L1-0 (orange line) versus atomic number. Shaded bands around the lines give the maximum and minimum yield or the calculations outlined in the previous section. Hereby we include all variations with the exception of the calculations with no included neutrino reactions. 
Black squares indicate abundances of the r-process enhanced star HD 222925 \citep[][]{Roederer2022}, magenta circles the star HD 122563 from \citet[][]{Honda2007} with additional values of \citet[][]{Roederer2012} indicated by magenta diamonds. Cyan stars and brown triangles indicate the stars CS22897-008 from \citet[][]{Spite2018} and HE0109-4510 from \citet[][]{Hansen2015}, respectively. All stars are normalized to the abundance of Sr ($Z=38$) of model L1-0. Upper limits are indicated with a downwards pointing arrow. The background colours indicate different percentages of observations (out of a total of $\sim 19500$ stars) as shown in the upper panel: green indicates more than $5\%$, blue more than $1\%$, orange more than one observation, red exactly one observation, and white no observation at all. This gives an estimate on how easy it is to observe a certain element. The arrows at the bottom of the lower panel indicate the regions of the first, second, and third r-process peak.} 
\label{fig:abundances_obs}
\end{figure}

An interesting result from MR-SNe nucleosynthesis is the variability of both: abundance pattern of lighter heavy elements and relative abundance between lighter and heavy r-process elements. It is well known that the production of heavy elements ($A>100$) in supernova explosions displays a (possibly non-monotonic) dependence on the magnetic field strength \citep[][]{Nishimura2015,Nishimura2017,Reichert2021a,Reichert2023}.
Studying the impact on the nucleosynthesis due to a variation of the magnetic field strength is motivated by the recognition that the strength of the magnetic field can undergo variation and amplification due to processes such as the magnetorotational instability \citep{reboul-salze2021,reboul-salze2022} or PNS convection \citep{Raynaud2020}. Here, we have shown that also the topology of the magnetic field leads to a substantial scatter in the abundance pattern. In nature, there most likely exist a variety of magnetic field strengths and magnetic field topologies. Even for the same stellar progenitor (i.e. same mass, initial composition and rotational properties), stellar evolution models predict a variety of pre-supernova end-points. The source of that variability is largely owed to the one-dimensional modelling of the secular stellar evolution, which necessarily includes approximations for intrinsically multidimensional phenomenae, e.g., convection, turbulence, the transport of angular momentum, the development of MHD instabilities, etc. But even a (nowadays impractical) multidimensional modelling would hardly be fully deterministic regarding the properties of the magnetic field, since some physics may induce a stochastic variance in the results (for instance, the turbulence). Considering various possible realizations of the magnetic field strength and topology of the same stellar progenitor of MR-SNe is useful to understand how much of the observed scattering in the distribution of abundances in r-process enriched stars could be attributed to that particular property of massive stars. Naturally, there are other sources of potential scattering in the nucleosynthetic yields, among which the various masses and metallicities of the stellar progenitors can be of utmost importance. Here we find that the scattering of abundance patterns in r-process enriched stars can be fully covered up to the second r-process peak by the degree with which the initial magnetic field resembles that of an aligned dipole. Beyond the second r-process peak, topological changes in the poloidal magnetic field may only account for the observed abundances of moderately r-process enriched stars up to $Z\sim 75$ (neglecting the potential for a delayed formation of a collapsar).

\section{Conclusions and discussions}
\label{sct:conclusion}
We investigated the nucleosynthesis of five state-of-the-art 3D CC-SNe models with rapid rotation. One model did not host any magnetic field and was calculated as a reference, while the four other models hosted the same strong magnetic field strength, but different topology. Considering variations in the magnetic topology aims at addressing the uncertainties that surround the astrophysical origin of such fields. They have to do with the approximations necessary to obtain a model of a massive star at the brink of collapse, foremost of which is the one-dimensional treatment of stellar evolution. 
Moreover, the action of dynamo processes within the central PNS during the onset of the explosion can lead to the amplification of magnetic fields up to magnetar-like strength with complex spatial structure and time variability \citep{Raynaud2020,reboul-salze2022}.
All these factors make the field topology very uncertain. For the first time, we investigated the nucleosynthetic yields of models that have a magnetic field configuration not dominated by a dipole. Among the topologies explored, we include, quadrupolar fields aligned with the rotational axis, and a dipolar field perpendicular to the rotational axis. We obtain a large variety of possible ejecta compositions leading from compositions dominated by iron for the model without magnetic fields (model H), to the models with a quadrupolar field and a tilted dipole that eject elements of the second r-process peak ($A\sim 130$), to the most optimistic case of a large scale dipole that hosts elements up to the third r-process peak, even though in a reduced amount ($M(A\ge180) \sim 6\times 10^{-5}\, \mathrm{M}_\odot$).

In comparison to a similar model (model S) whose nucleosynthesis was investigated in \citet[][]{Reichert2023}, the here presented large-scale dipole model L1-0 hosts slightly more neutron-rich conditions. This difference is visible in an increased ejecta mass of elements heavier than the second r-process peak. The only difference between these two models is the initial toroidal magnetic field (and therefore the imposed total magnetic energy). This shows that even if toroidal fields are created during the 3D dynamical evolution post-collapse \cite{Bugli2021}, the nucleosynthetic yields can be sensitive to their strength in the pre-supernova (initial) model. In other words, the dynamical evolution of the explosion does not (totally) erase the initial conditions at the brink of collapse. Our comparison suggests that pre-collapse toroidal fields can affect not only the overall abundance of r-process nuclei, but also the amount of unstable nuclei such as $^{56}$Ni. This finding can explain previously reported low $^{56}$Ni yields within MR-SNe \citep[e.g.,][]{Nishimura2017}. 

In our models of magnetorotationally driven supernovae (MR-SNe) we observe conditions that are moderately neutron-rich. As a result, the nucleosynthetic path follows a trajectory within a region where nuclear masses have either already been experimentally measured or are on the cusp of being measured. This observation is encouraging, as it suggests that future experiments hold the potential to further reduce uncertainties in nuclear physics.

We systematically explored the sensitivity of abundance patterns using seven distinct nuclear mass models. Our models of MR-SNe depict moderately neutron-rich conditions, placing the nucleosynthetic path within a region where nuclear masses have already been experimentally measured or are poised for measurement. This prospect is encouraging, as it suggests that upcoming experiments have the potential to further mitigate uncertainties in nuclear physics.

We have also shown that different $\beta^{-}$-decay rates have only a minor impact on the final composition. In total, we tested the effect of three different sets of $\beta^{-}$-decays and $\beta^{-}$-delayed neutron emission probabilities from \citet[][]{Moeller2003}, \citet[][]{Moeller2019}, and \citet[][]{Marketin2016}. While the different probabilities for $\beta^{-}$-delayed neutrons did not change the ejecta composition, the different half-lives impacted the models with a quadrupolar field and a 90-degree tilted dipole. Within these models, the ejecta composed of elements with mass numbers of $90<A<130$ get slightly enhanced. This enhancement is caused by the differing half-lives of nuclei with \mbox{$51 < N < 59$}. Compared to \citet[][]{Moeller2003}, the half-lives of the rate tables of \citet[][]{Moeller2019} and \citet[][]{Marketin2016} are shorter in this region, which allows for a quicker development of the nucleosynthetic flow that ultimately leads to an increase of heavier elements in the ejecta. 

When investigating the impact of neutrino reactions, it has to be kept in mind that our post-processing procedure starts when the trajectories fall below $7\,\mathrm{GK}$. At this temperature, the network starts with the electron fraction given by the simulation, which already contains a certain treatment of neutrino reactions. Nevertheless, we were able to show that charged-current neutrino reactions on heavier nuclei (i.e., not nucleons) can increase the ejecta of elements with $A>90$ in the case where no magnetic fields are present (model H).

Our models lack the necessary neutron richness to facilitate the synthesis of elements undergoing fission. Consequently, the considerable uncertainties associated with the calculation of fission reactions and fragment distributions do not influence any of our results.

When including feedback of the nuclear energy to the temperature (what we have referred to as self-heating), we only observed a minor dependence of the nuclear energy on the ejected yields for all our models. This underlines that the magnetohydrodynamic simulation already assembles the most important ingredients for determining the temperature. 

Another source of uncertainty is of numerical origin. Longer post-collapse evolution may foster conditions for the generation of heavy elements, e.g., significantly increasing the entropy in the jetted ejecta \citep{Reichert2023}. Intrinsic to the discretization of the problem to make it amenable to numerical treatment, are the effects (numerically) produced by dynamics happening at scales of about the grid resolution. Although our models rely on very high order numerical methods, a higher resolution will be needed in the future. That will help to disentangle whether only changes in the initial poloidal magnetic field may allow for a robust production of r-process nuclei. The alternative possibility is that dynamo processes in the post-collapse evolution very rapidly create strong enough fields to produce a similar effect. The latter may demand significantly larger numerical resolution than we can currently afford. 

When comparing the ejecta composition with the abundances in the stellar atmosphere of old r-process enriched stars, it becomes clear that even our most optimistic model L1-0 is not agreeing well with strongly r-process enriched stars. However, it may be a candidate to explain less r-process enriched stars, so called Honda-type stars. 
Furthermore, irrespective of variations in progenitor masses or metallicities, the diversity of ejecta compositions arising from different realizations of magnetic field strength and topology within a single pre-supernova model offers a plausible account for abundance variations preceding the second r-process peak. The observed variation in lighter heavy elements, extending up to strontium or silver, has been previously highlighted in the literature (LEPP). Here, we demonstrated that MR-SNe could naturally offer a viable explanation for the presence and scatter of lighter heavy elements in the early Universe.

The large variety of yields obtained here may also pose a major challenge to Galactic chemical evolution models. Our results indicate that these calculations, if considering MR-SNe as sources for r-process elements, should consider distributions of magnetic field strengths, magnetic field configurations, and magnetic field orientations. The yields are therefore not only dependent on the mass and metallicity of the progenitor (i.e., pre-supernova) star, but also on the details of the magnetic field. This should be incorporated into Galactic chemical evolution models in order to accurately describe the evolution of heavy elements in the early universe. This may not be feasible in the near future, since also not enough theoretical models and nucleosynthetic yields exist yet.

In this paper, we have taken another step to provide theoretical constrains that distinguish magnetorotational from ordinary SNe or neutron-star mergers as sources of heavy elements in the Universe. The uncertainties in the models (of various origins) provide a variegated landscape of abundance patters. That motivates us to continue our endeavour of producing further and more sophisticated numerical models whose observational signatures (nucleosynthesis patterns are one of them) permit deciphering whether observed SNe are driven magnetorotationally or otherwise.

\section*{Acknowledgements}
We thank D. Martin for providing rate tables of neutron capture reactions. Furthermore, we want to thank A. Tolosa-Delgado for useful discussions. MR acknowledges support from the Juan de la Cierva programme (FJC2021-046688-I). MO acknowledges support from the Spanish Ministry of Science, Innovation and Universities via the Ramón y Cajal programme (RYC2018-024938-I). Furthermore, MR, MAA, and MO acknowledge support from grant PID2021-127495NB-I00, funded by MCIN/AEI/10.13039/501100011033 and by the European Union “NextGenerationEU" as well as “ESF Investing in your future”. Additionally, they acknowledge support from the Astrophysics and High Energy Physics programme of the Generalitat Valenciana ASFAE/2022/026 funded by MCIN and the European Union NextGenerationEU (PRTR-C17.I1) as well as support from the Prometeo excellence programme grant CIPROM/2022/13 funded by the Generalitat Valenciana. MB and JG acknowledge support by the European Research Council (MagBURST grant 715368). A.A. acknowledges support by the Deutsche Forschungsgemeinschaft (DFG, German Research Foundation) -- Project-ID 279384907 - SFB 1245 and the State of Hessen within the Research Cluster ELEMENTS (Project ID 500/10.006). Part of the numerical calculations have been carried out at the TGCC on the Irene supercomputer and IDRIS on the Jean-Zay supercomputer (DARI projects A0090410317, A0110410317 and A0130410317). Furthermore we acknowledge support of computational resources on the HPC Lluisvives of the Servei d'Informàtica of the University of Valencia (financed by the FEDER funds for Scientific Infrastructures; IDIFEDER-2018-063). This project has received funding from the European Union's Horizon Europe research and innovation programme under the Marie Sk\l{}odowska-Curie grant agreement No 101064953 (GR-PLUTO).

\section*{Data Availability}
Produced data within this manuscript will be shared on reasonable request to the corresponding author.



\bibliographystyle{mnras}
\bibliography{example} 




\appendix


\bsp	
\label{lastpage}
\end{document}